\documentclass[lineno]{custom-arxiv}

\usepackage{caption} 
\usepackage{newtxtext}
\usepackage[subscriptcorrection]{newtxmath}

\makeatletter

\usepackage{ifthen}
\usepackage{pstricks}
\usepackage{subcaption}
\usepackage{enumitem}
\usepackage{ulem} 
\usepackage{verbatim}
\usepackage{multirow}
\usepackage{tabularx}
\usepackage{url}

\usepackage{xcolor,colortbl}

\usepackage{amsmath,amsfonts}
\theoremstyle{plain}

\theoremstyle{remark}
\newtheorem{result}{Result}

\usepackage{xr-hyper}
\usepackage{hyperref}
\begin{document}



\markboth{Hopkins \textit{et. al}}{Nonparametric Estimation of Ratios of Means}

\title{\vspace{-1cm}Nonparametric Identification and Estimation of Ratios of Multi-Category Means under Preferential Sampling}


\author{Grant Hopkins}
\affil{Department of Biostatistics, University of Washington,\\ Seattle, WA, U.S.A.}

\author{Sarah Teichman}
\affil{Department of Biostatistics, University of Washington,\\ Seattle, WA, U.S.A.}

\author{Ellen Graham}
\affil{Department of Biostatistics, University of Washington,\\ Seattle, WA, U.S.A.}

\author{\and Amy D. Willis}
\affil{Department of Biostatistics, University of Washington,\\ Seattle, WA, U.S.A.
\email{adwillis@uw.edu}}

\maketitle

\newboolean{DEBUG}
\setboolean{DEBUG}{false}

\newrgbcolor{amycolor}{.5 .1 .99}
\ifthenelse {\boolean{DEBUG}}
{\newcommand{\amy}[1]{{\amycolor{[\@Amy: #1]}}}}
{\newcommand{\amy}[1]{}}

\newrgbcolor{mariacolor}{.392 .585 .93}
\ifthenelse {\boolean{DEBUG}}
{\newcommand{\maria}[1]{{\mariacolor{[\@María: #1]}}}}
{\newcommand{\maria}[1]{}}

\ifthenelse {\boolean{DEBUG}}
{\newcommand{\mariadraft}[1]{{\mariacolor{#1}}}}
{\newcommand{\mariadraft}[1]{#1}}

\begin{abstract}
Multi-category data arise in diverse fields including marketing, chemistry, public policy, genomics, political science, and ecology. We consider the problem of estimating ratios of category-specific means in a fully nonparametric setting, allowing for both observational units and categories to be preferentially sampled. We consider covariate-adjusted and unadjusted estimands that are non-parametrically defined and straightforward to interpret. While identifiability for related models has been established through parametric distributions or restrictions on the conditional mean (e.g., log-linearity), we show that identifiability can be obtained through an independence assumption or a category constraint, such as a reference category or a centering function. We develop an efficient, doubly-robust targeted minimum loss based estimator with excellent finite-sample performance, including in the setting of a large number of infrequently observed categories. We contrast the performance of our method with related approaches via simulation, and apply it to identify bacteria that are differentially abundant in diarrheal cases compared to controls. Our work provides a general framework for studying parameter identifiability in compositional data settings without requiring parametric assumptions on the data distribution. 
\end{abstract}

\section{Introduction} \label{sec:intro}




Multi-category outcomes, where each variable represents the observed level of a category, arise in many different fields, including operations research (e.g., the level of production for different goods in different factories), public health (e.g., the concentration of viral markers in sewage effluent), political science (e.g., the number of votes for different candidates across districts), geology (e.g., mass spectroscopy-derived measurements of the mineral composition of sediment cores) and behavioral science (e.g., the number of hours individuals allocate across different activities in their day). 

We consider the setting where multi-category outcome data is a compressed measurement of category levels that exist on an absolute scale. However, due to practical constraints, measurements on the absolute scale are not obtained. We study what can be learned about differences in the category levels on the absolute scale from data that is \textit{not} observed on this scale. 
We connect the observed data to the absolute levels via the assumption that the observed measurement is proportional (in expectation) to the absolute level, but with unknown sample-specific and category-specific scaling factors. 

In this setting, the identifiable parameters are multiplicative differences in means of the same category across levels of a binary variable. We refer to these ratios of means as \textit{fold-differences}. In contrast to commonly-used models for multi-category data, we establish identifiability of fold-differences without imposing parametric assumptions on the data distribution or the relationship between adjustment covariates and the form of the conditional mean. As a result, our parameter is an interpretable summary of the generative distribution on the absolute scale, rather than an artifact of a working model for the observed data. Our approach does not require pre-specification of a reference category, though a reference category or centering constraint can provide an alternative to assuming independence of the sampling effort and covariates. 

To estimate fold-differences in the covariate-adjusted setting, we apply the toolkit of targeted minimum loss \citep{vanderLaan.Rubin2006}, leveraging flexible machine learning methods commonly used in modern causal inference to estimate the nuisance parameters. Many applications of our approach are in the setting of many infrequently observed categories, and so we apply a specific de-biasing procedure with finite-sample advantages. 
Our resulting estimator is doubly-robust to nuisance misspecification, and achieves the asymptotic variance lower bound given by the efficient influence function. We present hypothesis tests and marginal and simultaneous confidence intervals, and evaluate both point and interval estimates via simulation in comparison to related methods. 

While estimating fold-differences in the nonparametric setting has applications in many scientific disciplines, our investigation was initially motivated by estimating differences in the abundances of species in a microbial community. Absolute abundances of microbial species are not observed from high-throughput sequencing, and high-throughput sequencing data displays both sample-specific \citep[Figure 1]{Roesch.etal2007} and category-specific scaling factors \citep[Figure 3]{McLaren.etal2019} relative to absolute abundances, consistent with our model. We therefore demonstrate the utility of our method by analyzing microbiome data, studying which bacterial species in a microbiome are present in greater abundance in diarrheal cases compared to controls. We also compare our approach to widely-used methods for estimating microbial differential abundance. 

Our approach is related to, but distinct from, the paradigm of compositional data analysis, which is typically motivated by proportion-valued or scale-invariant vectors \citep{Aitchison1982,Aitchison1986,Aitchison1992,Greenacre.etal2023}. 
In contrast to methodology motivated by data characteristics, we are motivated by a generative model that connects (desirable but unobserved) absolute-scale data to (noisily rescaled) observed data. Our approach is more closely related to  \cite{Firth.Sammut2023}, who consider classes of generative data models constrained only by the proportionality of the observed data mean to a true underlying proportion, scaled by multiplicative errors and sample-specific effects. 
The class of models that we consider partially overlaps with the classes of models of \citet{Firth.Sammut2023}, as we allow category-specific distortions in the observed data but require additional independence assumptions. We also consider a latent process that defines the target of inference, and propose a specific methodology for estimation. 
That said, zero-valued data pose no theoretical or practical problems for either framework, eliminating the need to transform raw data (e.g., to the proportion scale, to replace zero-valued observations \citep{Martin-Fernandez.etal2003}, via a log-ratio-based method \citep{Egozcue.etal2003}, or to equalize totals across samples \citep{McMurdie.Holmes2014}). 

The paper is structured as follows: We begin by defining our target parameter on the absolute scale in Section \ref{subsec:parameter_defn}, and describing the observed and unobserved random variables in Section \ref{subsec:data_distn}. We then introduce assumptions to connect the observed data to the target parameter in Section \ref{subsec:identifiability_assumptions}, and present identifiability results in Section \ref{subsec:identifiability_results}. 
We then derive $\sqrt{n}$-consistent estimates of target parameters that are efficient over a large, nonparametric class of generative distributions in Section \ref{sec:methods}. We evaluate the error of point estimates and coverage of interval estimates in Section \ref{sec:sims}, and present an application of our method to a case-control microbiome study in Section \ref{sec:data_analysis}. We conclude with the summary of our work, along with limitations and possible extensions in Section \ref{sec:discussion}.


\section{Data and parameters}

\subsection{Parameter definition} \label{subsec:parameter_defn}

Let $V_i \in \mathbb{R}_{\geq 0}^J$ denote the level of each of the $J$ possible categories in sample $i$. When referring to a generic observation, we drop the sample subscript $i$. In our setting, $V$ is not required to be observable, and can be considered on any scale that is consistent across categories. For example, in a microeconomics setting, $V$ could be total within-category spending (e.g., dollars spent by individuals on housing, groceries, leisure, etc.), or $V$ could be within-category spending per-week. 
Let $A_i \in \{0, 1\}$ denote the primary covariate of interest for sample $i$, which we refer to as the \textit{exposure}, though this variable could be any binary characteristic of the observational unit. Let $X_i \in  \mathcal{X} \subseteq \mathbb{R}^p$ denote additional covariates. We are interested in estimating the unadjusted log-fold difference in the average level of the categories, 
\begin{align}
    \Psi^{(1)}(\mathbb{P}) &:= \left( \Psi_1^{(1)}(\mathbb{P}), \ldots, \Psi_J^{(1)}(\mathbb{P})\right), \\
    \Psi_j^{(1)}(\mathbb{P}) &:= \log\left( \dfrac{\mathbb{E}\left[V_{ j}|A = 1\right]}{\mathbb{E}\left[V_{ j}|A = 0\right]} \right), \label{eq:defn_psi1} 
\end{align}
or the adjusted log-fold difference in the average level of the categories,
\begin{align}
    \Psi^{(2)}(\mathbb{P}) &:= \left( \Psi_1^{(2)}(\mathbb{P}), \ldots, \Psi_J^{(2)}(\mathbb{P})\right), \\
    \Psi_j^{(2)}(\mathbb{P}) &:= \log\left( \dfrac{\mathbb{E}\left[\mathbb{E}\left[V_{ j}|A = 1, X\right]\right]}{\mathbb{E}\left[\mathbb{E}\left[V_{ j}|A = 0, X\right]\right]} \right), \label{eq:defn_psi2}
\end{align}
where the outer integral is taken over the distribution of $X$. Throughout, we let $\log$ denote the natural logarithm. As the conditional mean $\mathbb{E}\left[V_{ j}|A = a, X = x\right]$ is a function of $x$, integrating over the distribution of $X$ targets a covariate-weighted average of a (possibly covariate-dependent) conditional mean. 
We discuss causal interpretations of these parameters in Section \ref{sec:discussion}. 
We assume throughout that both parameters are well-defined, or equivalently, that there is a non-zero probability that the conditional mean is strictly positive for some $x \in \mathcal{X}$. 
This restriction is weak, allowing for the latent mean to be zero for some combinations of categories and covariates. 
Defining these parameters on the log-ratio scale, instead of the ratio scale, imposes no restrictions but facilitates comparison with parameters targeted by related methods.


\subsection{Observed and latent variables}  \label{subsec:data_distn}

If the true levels $V$ were observed, we could proceed directly to estimation, applying standard methods for treatment effect estimation. 
However, we consider the setting in which we only have access to observed levels $W$ that are distorted by some scaling factor. Therefore, in order to provide a summary of the true levels, we must first characterize how $W$ and $V$ relate. 


Distorted measurements of multi-category data appear in many real-world applications. For example, mass spectrometry cannot measure total atomic counts of geochemical isotopes, only their relative proportions. Similarly, market share of different brands, rather than total sales, standardizes across different currencies and overall consumption levels across regional markets. 
To reflect these settings, we consider $V$ to be a latent, unobserved random variable. 
Instead, we observe a random variable $W_i \in \mathbb{R}_{\geq 0}^J$. As we will formalize in Section \ref{subsec:identifiability_assumptions}, in addition to random noise, $W$ differs from $V$ in two ways. Firstly, its overall scale may differ. We consider scale to be a latent random variable, and denote it by $S \in \mathbb{R}_{> 0}$. In addition, the $J$ categories may differ in their observability, with certain categories more easily sampled on average. We denote the (latent) efficiency of observability by $E \in \mathbb{R}_{>0}^J$. 
We assume that data are generated independently and identically according to an unknown distribution $\mathbb{P}$: 
\begin{align}
    Z := (\overbrace{W, A, X}^\text{Observed}, \overbrace{V, E, S}^\text{Unobserved}) \overset{iid}{\sim} \mathbb{P}. \label{eq:data_distn}
\end{align}
We refer to $V_{ij}$ as the \textit{true} level of category $j$ and refer to $W_{ij}$ as the \textit{observed} level of category $j$.
One contribution of this paper is to propose sets of non-parametric assumptions on $\mathbb{P}$ that allow identification of $\Psi_j^{(1)}(\mathbb{P})$ and $\Psi_j^{(2)}(\mathbb{P})$. 

\subsection{Identifiability assumptions} \label{subsec:identifiability_assumptions}

In order to learn about summaries $\Psi_j^{(1)}(\mathbb{P})$ and $\Psi_j^{(2)}(\mathbb{P})$, we must restrict the model $\mathbb{P}$. These restrictions connect the observed data $(W, A, X)$ to the conditional distribution of $V$'s, through which $\Psi_j^{(1)}(\mathbb{P})$ and $\Psi_j^{(2)}(\mathbb{P})$ are defined. 
In this section, we describe some identifying assumptions assumptions that restrict $\mathbb{P}$. In Section \ref{subsec:identifiability_results}, we present combinations of these assumptions that allow us to identify our target parameters. Not all assumptions described in this section are required to estimate $\Psi_j^{(1)}(\mathbb{P})$ and $\Psi_j^{(2)}(\mathbb{P})$. 

Our first assumption connects, in expectation, the observed category levels $W$ to the latent category levels $V$: 

\begin{assumption} \label{assumption:mean_w}
$\mathbb{E}[W_{ j}|A,X] = \mathbb{E}[V_{ j} E_{ j} S|A,X]$  for all $j \in \{1, \dots, J\}$
\end{assumption}

Assumption \ref{assumption:mean_w} clarifies the interpretation of the scale and observability latent variables. Samples with larger values of $S$ have larger observed levels across all categories in expectation (all else equal), and categories with larger $E_j$ values have larger observations than those with smaller $E_j$ values (all else equal). We do not require that the observability of a category is strictly constant; it can randomly vary across samples provided that the central tendency of Assumption \ref{assumption:mean_w} remains true. 
Of course, $E_j = E_{j'}$ for all $j, j'$ (no overdetection) and degenerate $S$ (constant sampling intensity) are specific cases of this model. 


\begin{assumption} \label{assumption:joint_indep}
$(V_{ j} \perp E_{ j} \perp S) | (A, X)$ for all $j$
\end{assumption}

Assumption \ref{assumption:joint_indep} asserts that the true category levels, observabilities, and scales arise for unrelated reasons within subgroups defined by covariates. This precludes individuals with greater true levels of a category from also having that category as easier (or more difficult) to detect. For example, food frequency questionaires from individuals with high consumption of the category ``sugar'' cannot underreport their sugar consumption on average. 


\begin{assumption} \label{assumption:e_indep}
$E_{ j} \perp (A, X)$ for all $j$
\end{assumption}

By requiring that observabilities cannnot depend on exposure or covariates, Assumption \ref{assumption:e_indep} excludes categories from being systematically easier to observe in $A=1$ samples compared to $A=0$ samples. 

\begin{assumption} \label{assumption:s_joint_indep}
$S \perp (A, X)$
\end{assumption}

\begin{assumption} \label{assumption:s_cond_indep}
$(S \perp X) | A$
\end{assumption}

Assumptions \ref{assumption:s_joint_indep} and \ref{assumption:s_cond_indep} concern the relationships between sample effects $S$ and the covariates. Assumption \ref{assumption:s_joint_indep} asserts that the distribution of the scaling effects are the same across covariate groups, while Assumption \ref{assumption:s_cond_indep} asserts that scaling effects cannot differ in distribution across covariate levels, but may differ across exposure group. 



\subsection{Identifiability results} \label{subsec:identifiability_results}

We now present sets of assumptions that allow us to identify the parameters $\Psi_j^{(1)}(\mathbb{P})$, $\Psi_j^{(2)}(\mathbb{P})$, and related variations. 
Throughout, we let $g$ denote a differentiable function that satisfies $g(\mathbf{y} + z\mathbf{1}_J) = g(\mathbf{y}) + cz$ for any $\mathbf{y}\in\mathbb{R}^J$ and $z \in \mathbb{R}$, and fixed $c \neq 0$, where $\mathbf{1}_J$ is a $J$-dimensional vector of $1$'s. For example, 
$g(\mathbf{y}) = y_J$ and $g(y_1, \dots, y_J) := \frac{1}{J} \sum_{j=1}^J y_j$ both satisfy this requirement. 
As our identifiability assumptions imply that ratios of means across categories are  distorted by a common multiplicative scaling factor, $g$'s of this form allow recentering estimates to eliminate their dependence on that factor.  
Proofs of all results can be found in SI Section 2. 

\begin{result}
\label{res:unadjusted_uncentered}
    Under Assumptions \ref{assumption:mean_w}, \ref{assumption:joint_indep}, \ref{assumption:e_indep} and \ref{assumption:s_joint_indep}, 
    $\Psi_j^{(1)}(\mathbb{P})$ is identifiable for all $j$.
\end{result}

\begin{result}
\label{res:unadjusted_centered}
    Under Assumptions \ref{assumption:mean_w}, \ref{assumption:joint_indep} and \ref{assumption:e_indep}, the following is identifiable for all $j$: 
    \begin{align*}
        \Psi_j^{(1g)}(\mathbb{P}) := \Psi_j^{(1)}(\mathbb{P}) - g\left(\Psi_1^{(1)}(\mathbb{P}),\dots,\Psi_J^{(1)}(\mathbb{P})\right),
    \end{align*}
\end{result}

Result \ref{res:unadjusted_centered} accords with the common practice in the compositional data literature of modeling differences across categories defined by covariates relative to a reference category. For example, the coefficient on $A$ in certain multinomial generalized linear is exactly equal to $\Psi_j^{(1g)}(\mathbb{P})$ for $g(\mathbf{y})$ that returns the reference category (see SI Section 1.1 for details).

We now discuss adjusted parameters, demonstrating that under the setting of \eqref{eq:data_distn} and the conditional mean and independence assumptions discussed above, identifiability of ratios of means of the true response levels can be obtained either through a joint independence assumption, or through a conditional independence assumption combined with centering.

\begin{result}
\label{res:adjusted_uncentered}
    Under Assumptions \ref{assumption:mean_w}, \ref{assumption:joint_indep}, \ref{assumption:e_indep} and \ref{assumption:s_joint_indep}, 
    $\Psi_j^{(2)}(\mathbb{P})$ is identifiable for all $j$. 
\end{result}

\begin{result} \label{result:identifiability_2g}
\label{res:adjusted_centered}
    Under Assumptions \ref{assumption:mean_w}, \ref{assumption:joint_indep}, \ref{assumption:e_indep} and \ref{assumption:s_cond_indep}, the following parameter is identifiable for all $j$: 
    \begin{align*}
        \Psi_j^{(2g)}(\mathbb{P}) := \Psi_j^{(2)}(\mathbb{P}) - g\left(\Psi_1^{(2)}(\mathbb{P}),\dots,\Psi_J^{(2)}(\mathbb{P})\right).
    \end{align*}
\end{result}

In contrast to the unadjusted case, an independence assumption about the sampling intensities $S$ is required to identify both $\Psi_j^{(2)}(\mathbb{P})$ and $\Psi_j^{(2g)}(\mathbb{P})$. The stronger assumption (independence of sampling effects and the joint distribution of exposure and covariates) allows for the stronger identifiability result, but a centered version of the parameter can still be identified provided that the sampling effects are independent of the covariates conditional on the exposure. 
Also in contrast to the unadjusted setting, the coefficient on $A$ in multinomial generalized linear models will not equal $\Psi_j^{(2g)}(\mathbb{P})$ in general. However, these parameters are equal if the ratio of category probabilities to a reference category is truly log-linear under $\mathbb{P}$. See SI Section 1.2 for details and further discussion.

We conclude this section by noting that the choice of $g$ in Results \ref{res:unadjusted_centered} and \ref{res:adjusted_centered} alters the interpretation of the parameters $\Psi^{(1g)}$ and $\Psi^{(2g)}$. While choosing a reference category $g(\mathbf{y}) = y_J$ is common in many compositional data tools  and distributions for multi-category data (e.g., the multivariate normal model for additive log-ratio transformed compositions \cite[Chapter 7]{Aitchison1986} and multinomial GLMs \cite[Chapter 6]{Long1997}), other choices of constraints may be desirable in settings where no natural reference category exists. 
For example, the choice $g(y_1, \dots, y_J) := \frac{1}{J} \sum_{j=1}^J y_j$ targets the log-ratios of means (across exposures) relative to the average log-ratio across categories, with a natural connection to the centered log-ratio transformation \cite[Definition 4.6]{Aitchison1986}. 
\cite{Clausen.Willis2024}, who studied identifiability and estimation in a log-linear model under an assumption similar to Assumption \ref{assumption:mean_w}, argue that choosing $g$ to be a smooth approximation to the median preserves interpretability of location-centering while mitigating the potential for large ratios to alter the overall target parameter. 
Of course, as Results \ref{res:unadjusted_uncentered} and \ref{res:adjusted_uncentered}
show, identifiability can be obtained without centering.

\section{Estimation} \label{sec:methods}

Having defined our parameters of interest and conditions under which these parameters are identifiable, we now discuss their estimation. When no covariate adjustment is needed, a simple plug-in estimator achieves the asymptotic lower bound on variance for a consistent, regular estimator. For parameters with covariate adjustment, we consider two approaches to efficient estimation: 1) one-step estimation (OSE) and 2) targeted minimum loss based estimation (TMLE). For both practical and theoretical reasons, TMLE outperforms OSE here. 
Proofs can be found in SI Section 7. 

\subsection{Influence functions and efficiency} \label{subsubsec:inf_fn_preliminaries}

Before discussing estimation, we introduce key concepts from semi-parametric efficiency theory, and refer the reader to \citet[Chapter 25.3]{vdv} for further information. 
In what follows, we derive the efficient influence function for a parameter $\Psi(\mathbb{P})$ over model $\mathcal{M}$, where $\mathcal{M}$ is a set of probability distributions on the observed data. 
The model should reflect all information we know about the data generating mechanism. A \textit{non-parametric model} makes minimal assumptions and is effectively a set of arbitrary distributions. 
Influence functions represent functional derivatives for smooth parameters, characterizing how the parameter changes for small changes in the underlying probability distribution.

The variance of the \textit{efficient} influence function gives the lower bound on the asymptotic variance that any consistent, regular estimator of $\Psi(\mathbb{P})$ can attain over model $\mathcal{M}$. While a formal definition can be found in \citet[Chapter 25.3]{vdv}, a regular estimator is one whose asymptotic distribution is not sensitive to ``local perturbations'' of $\mathbb{P}$. 
Regularity ensures that the estimator, if applied to similar distributions, will behave similarly. If $\phi(\mathbb{P}) : (\mathcal{W}, \mathcal{A}, \mathcal{X}) \mapsto \mathbb{R}^J$ is the efficient influence function, then all regular estimators must have asymptotic variance greater than or equal to \begin{align}
    \Sigma 
    & \:= \mathbb{E}_{\mathbb{P}}[\phi(\mathbb{P})(W,A,X) \times \phi(\mathbb{P})(W,A,X)^T], \label{eq:variance_bound}
\end{align} in the sense that $c^T \Sigma c$ is the smallest attainable variance for an estimator of the linear contrast $c^T \Psi(\mathbb{P})$ for all $c \in \mathbb{R}^J$. In a non-parametric model $\mathcal{M}$, there is only one influence function, which is necessarily efficient. In addition to giving the 
efficiency bound, the efficient influence function can be used to construct an 
estimator that achieves this bound. 
We proceed to derive the efficient influence functions for parameters $\Psi^{(1)}(\mathbb{P})$, $\Psi^{(1g)}(\mathbb{P})$, $\Psi^{(2)}(\mathbb{P})$, and $\Psi^{(2g)}(\mathbb{P})$ in non-parametric model $\mathcal{M}$ and use them to construct efficient estimators. We will state the influence functions in the main text; see derivations in SI Section 3.

\subsection{Parameters without covariate adjustment}

While $\Psi^{(1)}(\mathbb{P})$ is defined in terms of the distribution of the latent levels $V$, under Assumptions \ref{assumption:mean_w}, \ref{assumption:joint_indep}, \ref{assumption:e_indep}, and \ref{assumption:s_joint_indep}, we can write it in terms of the distribution of the observed levels $W$:
\begin{align}
    \Psi_j^{(1)}(\mathbb{P}) &=  \log\left( \dfrac{\mathbb{E}_{\mathbb{P}}\left[W_{ j}|A = 1\right]}{\mathbb{E}_{\mathbb{P}}\left[W_{ j}|A = 0\right]} \right).
\end{align}

The efficient influence function of $\Psi^{(1)}(\mathbb{P})$ is
\begin{align*}
    & \phi^{(1)}(\mathbb{P}) : (w,a,x) \\
    & \qquad \mapsto \begin{pmatrix}
    \dfrac{\frac{a}{\mathbb{P}(A=1)} \left( w_1 - \mathbb{E}_{\mathbb{P}}[W_1|A=1]\right)}{\mathbb{E}_{\mathbb{P}}[W_1|A=1]} - \dfrac{\frac{1 - a}{1-\mathbb{P}(A=1)} \left( w_1 - \mathbb{E}_{\mathbb{P}}[W_1|A=0]\right)}{\mathbb{E}_{\mathbb{P}}[W_1|A=0]} \\
    \vdots \\
    \dfrac{\frac{a}{\mathbb{P}(A=1)} \left( w_J - \mathbb{E}_{\mathbb{P}}[W_J|A=1]\right)}{\mathbb{E}_{\mathbb{P}}[W_J|A=1]} - \dfrac{\frac{1 - a}{1-\mathbb{P}(A=1)} \left( w_J - \mathbb{E}_{\mathbb{P}}[W_J|A=0]\right)}{\mathbb{E}_{\mathbb{P}}[W_J|A=0]}
\end{pmatrix}.
\end{align*}

We now consider the plug-in estimator $\widehat\psi^{(1)}_{\text{PI}}$ whose $j$th element is defined as
\begin{align} \label{eq:estimator_psi1}
\widehat\psi^{(1)}_{\text{PI},j} &:= \log \left( \dfrac{\frac{1}{|\{ i : A_i = 1 \}|}\sum\limits_{i : A_i = 1}{W_{ij}}}{\frac{1}{|\{ i : A_i = 0 \}|}\sum\limits_{i : A_i = 0}{W_{ij}}} \right).
\end{align}


The estimator above is simply a vector of the log-ratio of sample means between two subpopulations in each category of the multivariate response. This estimator is straightforward to implement and transparent in its identifiability assumptions. In particular, the above estimator makes no assumptions about the distribution of $(W,A)$, except that $\max\limits_{a\in\{0,1\}} \text{Var}_{\mathbb{P}}(W|A=a)$ is finite; this is trivial if $W$ has bounded support.

\begin{theorem} \label{result:efficiency1}
    Suppose $\mathbb{P}$ satisfies Assumptions \ref{assumption:mean_w}, \ref{assumption:joint_indep}, \ref{assumption:e_indep}, and \ref{assumption:s_joint_indep}.
    Then $\widehat\psi^{(1)}_{\text{PI}}$ is a regular and asymptotically linear estimator of $\Psi^{(1)}(\mathbb{P})$ with influence function $\phi^{(1)}(\mathbb{P})$. Therefore, $$\sqrt{n}\left(\widehat\psi^{(1)}_{\text{PI}} - \Psi^{(1)}(\mathbb{P})\right) \overset{d}{\rightarrow} \mathcal{N} \left( \textbf{0}_J, \Sigma^{(1)} \right)$$ for $\Sigma^{(1)} :=  \mathbb{E}_{\mathbb{P}}[\phi^{(1)}(\mathbb{P})(W,A,X) \times \phi^{(1)}(\mathbb{P})(W,A,X)^T]$. Thus, $\widehat\psi^{(1)}_{\text{PI}}$ is efficient in $\mathcal{M}$.
\end{theorem}

We now discuss the centered parameter $\Psi^{(1g)}(\mathbb{P})$, whose 
influence function is \begin{align*}
    & \phi^{(1g)}(\mathbb{P}) : (w,a,x) 
    \mapsto \begin{pmatrix}
    \mathbf{I}_{J} - \textbf{1}_J \times \nabla g \left(\Psi^{(1)}(\mathbb{P})\right)^T 
\end{pmatrix} \times \phi^{(1)}(\mathbb{P})(w,a,x)
\end{align*} where $\mathbf{I}_{J}$ is the $J$-dimensional identity matrix and $\textbf{1}_J \in \mathbb{R}^{J \times 1}$ is a vector of $1$'s. This parameter is a summary of the true category levels under different identifiability assumptions than Theorem \ref{result:efficiency1}. We apply the delta method and Theorem \ref{result:efficiency1} to obtain the following result about the estimator $\widehat\psi^{(1g)}_{\text{PI}} := \widehat\psi^{(1)}_{\text{PI}} - g\left( \widehat\psi^{(1)}_{\text{PI}} \right)$ of the centered parameter $\Psi^{(1g)}(\mathbb{P})$. 

\begin{theorem} \label{result:efficiency2}
    Suppose $\mathbb{P}$ satisfies Assumptions \ref{assumption:mean_w}, \ref{assumption:joint_indep}, and \ref{assumption:e_indep}.
    Then $\widehat\psi^{(1g)}_{\text{PI}}$ is a regular and asymptotically linear estimator of $\Psi^{(1g)}(\mathbb{P})$ with influence function $\phi^{(1g)}(\mathbb{P})$. Therefore, $$\sqrt{n}\left(\widehat\psi^{(1g)}_{\text{PI}} - \Psi^{(1g)}(\mathbb{P})\right) \overset{d}{\rightarrow} \mathcal{N} \left( \textbf{0}_J, \Sigma^{(1g)} \right)$$ for $\Sigma^{(1g)} :=  \mathbb{E}_{\mathbb{P}}[\phi^{(1g)}(\mathbb{P})(W,A,X) \times \phi^{(1g)}(\mathbb{P})(W,A,X)^T]$. Thus, $\widehat\psi^{(1g)}_{\text{PI}}$ is efficient in $\mathcal{M}$.
\end{theorem}

The unadjusted parameters $\Psi^{(1)}(\mathbb{P})$ and $\Psi^{(1g)}(\mathbb{P})$ will often be the target of inference when the exposure $A$ is assigned randomly. As we discuss in Section \ref{sec:discussion}, a causal interpretation can be ascribed to these parameters under standard causal assumptions. 

\subsection{Parameters with covariate adjustment}

We now consider estimation of covariate-adjusted parameters. 
Under Assumptions \ref{assumption:mean_w}, \ref{assumption:joint_indep}, \ref{assumption:e_indep}, and \ref{assumption:s_joint_indep}, we have
\begin{align*}
    \Psi_j^{(2)}(\mathbb{P})
    & = \log\left( \dfrac{\mathbb{E}_{\mathbb{P}}\left[\mathbb{E}_{\mathbb{P}}\left[W_{ j}|A = 1, X\right]\right]}{\mathbb{E}_{\mathbb{P}}\left[\mathbb{E}_{\mathbb{P}}\left[W_{ j}|A = 0, X\right]\right]} \right).
\end{align*}
We adopt the notation $\mu_{j}(a, x) := \mathbb{E}_{\mathbb{P}}\left[W_{ j}|A = a, X = x\right]$, $\pi(x) := \mathbb{P}(A = 1 | X = x )$, $F_X(x) := \mathbb{P}(X\le x).$
Let $\mathbb{E}_{\mathbb{P}}[f(X)] := \int_\mathcal{X}f(x)dF_X(x)$ for any function $f$, and $\mathbb{P}f :=  \mathbb{E}_{\mathbb{P}}[f(W,A,X)]$ and $\mathbb{P}_n f := \frac{1}{n}\sum\limits_{i=1}^{n}f(W_i,A_i,X_i)$.


The efficient influence function of $\Psi_j^{(2)}(\mathbb{P})$ is $\phi^{(2)}(w,a,x) = \left( \phi_1^{(2)}(w,a,x), \ldots, \phi_J^{(2)}(w,a,x) \right)$, where 
\begin{align}
    \label{eq:phi2if}
    \phi_j^{(2)}(\mathbb{P}) : (w,a,x) \mapsto \dfrac{\bar{\phi}_{1,j}(\mathbb{P})(w,a,x)}{\int_{\mathcal{X}}{\mu_{j}(1, z)} dF_X(z)} - \dfrac{\bar{\phi}_{0,j}(\mathbb{P})(w,a,x)}{\int_{\mathcal{X}}{\mu_{j}(0, z)} dF_X(z)},
\end{align}
where $\bar{\phi}_{1,j}(\mathbb{P})$ and $\bar{\phi}_{0,j}(\mathbb{P})$ are given by
\begin{align}
    \bar{\phi}_{1,j}(\mathbb{P}) : (w,a,x) & \mapsto \;\;\;\; \frac{a}{\pi(x)} \;\;\;\, \left( w_j - \mu_{j}(a, x)\right) + \mu_{j}(1, x) - \int_{\mathcal{X}}{\mu_{j}(1, z)} dF_X(z) \label{eq:if_g_comp1} \\
    \bar{\phi}_{0,j}(\mathbb{P}) : (w,a,x) & \mapsto \frac{1-a}{1-\pi(x)} \left( w_j - \mu_{j}(a, x)\right) + \mu_{j}(0, x) - \int_{\mathcal{X}}{\mu_{j}(0, z)} dF_X(z). \label{eq:if_g_comp0}
\end{align}

The influence function $\phi_j^{(2)}(\mathbb{P})(w,a,x)$ depends on $\mathbb{P}$ through the conditional mean $\mu_{j}$, propensity score $\pi$, and covariate distribution $F_X$. We refer to $\mu_{j}$, $\pi$ and $F_X$ as nuisance functions; these are unknown. Therefore, in order to use this influence function to construct an estimator, we need estimates $\hat{\mu}_j$, $\hat\pi$, and $\hat{F}_X$. We discuss estimation of these functions shortly, but for now let $\widehat{\mathbb{P}}$ denote a probability distribution with these estimated nuisances.

The first estimator of $\Psi^{(2)}(\mathbb{P})$ that we consider is $\widehat\psi^{(2)}_{\text{{PI}}}$, wherein $\widehat{\mathbb{P}}$ is plugged in for $\mathbb{P}$:
$$\widehat\psi^{(2)}_{\text{{PI}}, j} :=\Psi_j^{(2)}(\widehat{\mathbb{P}})
 = \log \left( \dfrac{\mathbb{E}_{\widehat{\mathbb{P}}}[\mathbb{E}_{\widehat{\mathbb{P}}}[W_j|A=1,X]]}{\mathbb{E}_{\widehat{\mathbb{P}}}[\mathbb{E}_{\widehat{\mathbb{P}}}[W_j|A=0,X]]} \right)  = \log \left( \dfrac{ \int_{\mathcal{X}} \hat{\mu}_j(1,x) d\hat{F}_X(x) }{ \int_{\mathcal{X}} \hat{\mu}_j(0,x) d\hat{F}_X(x) } \right)  = \log \left( \dfrac{\frac{1}{n}\sum_{i=1}^{n}\hat{\mu}_j(1,X_i)}{\frac{1}{n}\sum_{i=1}^{n}\hat{\mu}_j(0,X_i)} \right), $$
where the last equality holds if we choose $\hat{F}_X$ to be the empirical distribution function of the covariates, $\hat{F}_X(x) := \frac{1}{n}\sum_{i=1}^{n}\mathbf{1}(X_i \le x)$. 
In contrast to Section 4.1, where the plug-in estimator was efficient, $\widehat\psi^{(2)}_{\text{PI}}$ will perform poorly. When $X$ is continuous and flexible machine learning models are used to estimate $\mu_j$, $\widehat\psi^{(2)}_{\text{PI}}$ will inherit the excess bias of these machine learning models, leading to 
slower-than-$\sqrt{n}$-rates of convergence. One option to obtain to $\sqrt{n}$-rates of convergence is to use a parametric regression model for $\mu_j$, but 
if that model is misspecified (as will typically be the case in practice), then the estimator will not even be consistent. 
Alternatively, the efficient influence function can be used to "debias" $\widehat\psi^{(2)}_{\text{PI}}$ without specifying a parametric model for the nuisance parameters. This allows us to achieve $\sqrt{n}$-rate efficient estimation, even when we use flexible machine learning methods to estimate the nuisance functions. This approach is known as debiased (or double) machine learning (DML) \citep{chernozhukov2024doubledebiasedmachinelearningtreatment}. 
We pursue this strategy, and give conditions on nuisance estimation that guarantee our estimators will be consistent, regular, asymptotically linear and efficient. 

\textbf{One-step estimation} Our first approach to de-biasing $\widehat\psi^{(2)}_{\text{PI}}$ is one-step estimation \citep{bickel1993efficient}. The one-step estimator is given by
\begin{align}
    \widehat\psi^{(2)}_{\text{OS}^*, j}
    &= \Psi_j^{(2)}(\widehat{\mathbb{P}}) + \mathbb{P}_n\phi_j^{(2)}(\widehat{\mathbb{P}}) \label{eq:psi_adjusted_os} \\
    & = \widehat\psi^{(2)}_{\text{PI}, j} + \frac{1}{n}\sum\limits_{i=1}^{n}{\phi_j^{(2)}(\widehat{\mathbb{P}})}(W_i,A_i,X_i)  \notag \\
    &= \log\left( \dfrac{\frac{1}{n}\sum\limits_{i=1}^{n}\hat\mu_{j}(1, X_i)}{\frac{1}{n}\sum\limits_{i=1}^{n}\hat\mu_{j}(0,X_i)} \right) + \dfrac{1}{n} \sum\limits_{i=1}^{n} \left( \dfrac{\frac{A_i}{\hat\pi(X_i)} \left(W_{ij} - \hat\mu_{j}(1,X_i)\right)}{\frac{1}{n}\sum_{k=1}^{n}\hat\mu_{j}(1,X_k)} - \dfrac{\frac{1-A_i}{1 - \hat\pi(X_i)} \left(W_{ij} - \hat\mu_{j}(0,X_i)\right)}{\frac{1}{n}\sum_{k=1}^{n}\hat\mu_{j}(0, X_k)} \right). \notag
\end{align}
For technical reasons discussed in SI Section 3, we use $K$-fold cross-fitting \citep{schick1986asymptotically} to estimate the nuisance functions 
such that the data used to estimate each nuisance is independent of the data that they are evaluated on. Therefore, our actual estimator takes the form 
\begin{align} \label{eq:adj_estimator_cv}
    \widehat\psi^{(2)}_{\text{OS}, j} = \frac{1}{K} \sum_{k=1}^K \left( \Psi_j^{(2)}(\widehat{\mathbb{P}}_{-k}) + \mathbb{P}_{n,k}\phi_j^{(2)}(\widehat{\mathbb{P}}_{-k}) \right),
\end{align}
where $\mathbb{P}_{n,k}$ denotes the empirical distribution of the $k$th cross-fitting fold and $\widehat{\mathbb{P}}_{-k}$ denotes the nuisances estimated using all folds except fold $k$. We return to the details of nuisance estimation after introducing our second estimator.

\textbf{Targeted Minimum Loss Based Estimation}
Our second approach for de-biasing the plug-in estimator is TMLE \citep{vanderLaan.Rubin2006}. 
Whereas \eqref{eq:psi_adjusted_os} shows the one-step estimator to be an additive adjustment to the plug-in estimator,
TMLE adjusts the initial nuisance estimate 
$\widehat{\mathbb{P}} \mapsto \widehat{\mathbb{P}}^{\star}$ 
such that $\mathbb{P}_n\phi_j^{(2)}(\widehat{\mathbb{P}}^{\star}) = 0$. 
TMLE estimators therefore take the form $\widehat\psi^{(2)}_{\text{TMLE},\,j} := \Psi_j^{(2)}(\widehat{\mathbb{P}}^{\star}) + \mathbb{P}_n\phi_j^{(2)}(\widehat{\mathbb{P}}^{\star}) = \Psi_j^{(2)}(\widehat{\mathbb{P}}^{\star})$, illustrating that TMLE is both a plug-in \textit{and} a one-step estimator. 

We now describe how to find a $\widehat{\mathbb{P}}^{\star}$ such that $\mathbb{P}_n\phi_j^{(2)}(\widehat{\mathbb{P}}^{\star}) = 0$. We see from \eqref{eq:phi2if} that $\mathbb{P}_n \bar{\phi}_{1,j}(\widehat{\mathbb{P}}^{\star}) = \mathbb{P}_n \bar{\phi}_{0,j}(\widehat{\mathbb{P}}^{\star}) = 0$ is a sufficient condition for $\mathbb{P}_n \phi^{(2)}(\widehat{\mathbb{P}}^\star) = 0$. 
Therefore, we seek a $\widehat{\mathbb{P}}^{\star}$
such that for all $j\in\{1,\dots,J\}$ and $a \in \{0, 1\}$, $\widehat{\mathbb{P}}^{\star}$ satisfies
\begin{align*}
    0 &= \mathbb{P}_n \bar{\phi}_{a,j}(\widehat{\mathbb{P}}^{\star}) = \frac{1}{n}\sum\limits_{i=1}^{n}{\bar{\phi}_{a,j}(\widehat{\mathbb{P}}^{\star})}(W_i,A_i,X_i). 
\end{align*}
Fortunately, this is a well-studied problem, as 
$\bar{\phi}_{a,j}(\mathbb{P})$ is the efficient influence function of the G-computation parameter $\mathbb{E}\left[\mu_j(a,X)\right]$, first introduced by \cite{robins1986new}. 
Estimation of this parameter has been discussed extensively in causal inference literature, from which we take our solution.
For demonstration, we study the case of $a=1$ and arbitrary $j$. Choosing $\hat{F}_X ^{\star} = \hat{F}_X$, we have
\begin{align}
    0 & = \frac{1}{n} \sum\limits_{i=1}^{n}\left\{ \frac{A_i}{\hat{\pi}^{\star}(X_i)}\left( W_{ij} - \hat{\mu}^{\star}_{j}(1, X_i)\right) + \hat{\mu}^{\star}_{j}(1, X_i) - \int_{\mathcal{X}} \hat{\mu}^{\star}_{j}(1, z) d\hat{F}_X ^{\star}(z) \right\} \\
    & = \frac{1}{n} \sum\limits_{i=1}^{n}\left\{ \frac{A_i}{\hat{\pi}^{\star}(X_i)}\left( W_{ij} - \hat{\mu}^{\star}_{j}(1, X_i)\right) \right\}.
    \intertext{One $\widehat{\mathbb{P}}^{\star}$ that will solve this equation is achieved by scaling the conditional mean by a positive constant, $\hat{\mu}^{\star}_j : (1,x) \mapsto \exp(\beta_{1,j}) \times \hat{\mu}_j(1,x)$, while leaving the propensity score unaltered, $\hat{\pi}^{\star} : x \mapsto \hat{\pi}(x)$ \citep{van2011targeted}. This approach is the solution in $\beta_{1,j} \in \mathbb{R}$ to}
    0 & = \frac{1}{n} \sum\limits_{i=1}^{n}\left\{ \frac{A_i}{\hat{\pi}(X_i)}\left( W_{ij} - \exp(\beta_{1,j}) \times \hat{\mu}_{j}(1,X_i)\right) \right\},  \label{eq:tmle_beta}
\end{align} 
which is equivalent to the score equation for a intercept-only Poisson regression with response $W_{ij}$, weights $A_i/\hat{\pi}(X_i)$, and offset $\text{log}\left( \hat{\mu}_{j}(1, X_i) \right)$. In TMLE literature, these transformed variables are sometimes referred to as \textit{clever-covariates} \citep{vanderLaan.Rubin2006}. Hence, for any initial nuisances $\hat{\mu}_j$ and $\hat{\pi}$, we can use off-the-shelf GLM solvers to find $\exp(\beta_{1,j})$ for all $j \in \{1,\dots,J\}$. We can repeat this process for the $a=0$ case to find the analogous $\exp(\beta_{0,j})$ for all $j \in \{1,\dots,J\}$. 
Taken together, if we define $\widehat{\mathbb{P}}^{\star}$ such that $\hat{\mu}_j^\star(a,x)  := \exp(\beta_{a,j}) \hat{\mu}_j(a,x)$ for all $j \in \{1,\dots,J\}$, $\hat{\pi}^\star(x) := \hat{\pi}(x) $ and $\hat{F}_X^\star(x) := \frac{1}{n}\sum_{i=1}^{n}\mathbf{1}(X_i \le x),$
then we have that $\mathbb{P}_n \bar{\phi}_{a,j}(\widehat{\mathbb{P}}^{\star}) = 0$, yielding the estimator 
$\widehat\psi^{(2)}_{\text{TMLE}^*,\,j} = \Psi_j^{(2)}(\widehat{\mathbb{P}}^{\star})$.

The above procedure describes the classical approach to de-bias a plug-in estimator of $\mathbb{E}\left[\mu_j(u,X)\right]$ via TMLE for non-negative responses. 
However, it does not address the common challenge that a large proportion of responses $W_{ij}$ equal zero. We therefore take the approach of \cite{williams2024two}, who describe a TMLE variant with practical advantages in this setting. When the outcome variable is non-negative, the conditional mean $\mu_{j}$ is the product 
$\mu_{j}(a,x) = \mathbb{E}[W_j|W_j>0,A=a,X=x] \mathbb{P}(W_j>0|A=a,X=x) =: m_{j}(a,x) q_{j}(a,x).$ 
Therefore, we apply a two-stage procedure to separately de-bias 1) the conditional mean among positive observations and 2) the conditional proportion of zero observations. This procedure offers finite-sample advantages, in part because many machine learning methods to flexibly estimate $\mu_j$ perform poorly when a non-negative outcome has many zeros.

Finally, 
we use $K$-fold cross-fitting to construct the initial nuisance estimates. We replace $\hat{\pi}$ and $\hat{\mu}_j$ in \eqref{eq:tmle_beta} with $\hat{\pi}_{-k_i}$ and $\hat{\mu}_{-k_i,j}$ to solve $\tilde{\beta}_{1,j}$, where the subscript $-k_i$ indicates that the function is estimated using all folds except the fold containing data point $i$ (and similarly for $\tilde{\beta}_{0,j}$). 
This gives our final estimator
\begin{align*}
    \widehat\psi^{(2)}_{\text{TMLE},\,j} &= \log\left( \dfrac{\frac{1}{n} \sum\limits_{i=1}^{n}  \exp(\tilde{\beta}_{1,j}) \hat{\mu}_{-k_i,j}(1,X_i)}{\frac{1}{n} \sum\limits_{i=1}^{n} \exp(\tilde{\beta}_{0,j}) \hat{\mu}_{-k_i,j}(0,X_i)} \right).
\end{align*} 

\subsection{Asymptotic linearity and efficiency of estimators}


As previously mentioned, the plug-in estimator $\widehat\psi^{(2)}_{\text{PI}}$ will not be consistent unless each conditional mean is estimated well, that is, $\|\hat{\mu}_j - \mu_j\|_{L_2(\mathbb{P})} = o_{\mathbb{P}}(1)$. Furthermore, it will not have a normal limit unless $\|\hat{\mu}_j - \mu_j\|_{L_2(\mathbb{P})} = O_{\mathbb{P}}(n^{-1/2})$. 
Satisfying the latter condition requires $\hat{\mu}_j$ to be a correctly specified parametric regression model, which is both implausible and contradictory to our non-parametric model.

In contrast, the one-step and TMLE estimators are consistent and asymptotically normally distributed under weaker conditions. These weaker conditions can be plausibly satisfied using flexible machine learning methods for nuisance estimation \citep{chernozhukov2024doubledebiasedmachinelearningtreatment, rotnitzky2021characterization}. The one-step estimator $\widehat\psi^{(2)}_{\text{OS}, j}$ requires $\|\hat{\mu}_j - \mu_j\|_{L_2(\mathbb{P})} = o_{\mathbb{P}}(1)$ for consistency, and both $\|\hat{\mu}_j - \mu_j\|_{L_2(\mathbb{P})} = o_{\mathbb{P}}(n^{-1/4})$ and $\|\hat{\mu}_j - \mu_j\|_{L_2(\mathbb{P})} \times \|\hat{\pi} - \pi\|_{L_2(\mathbb{P})} = o_{\mathbb{P}}(n^{-1/2})$ for asymptotic normality. The 
TMLE estimator $\widehat\psi^{(2)}_{\text{TMLE},\,j}$ requires even weaker conditions: either  $\|\hat{\mu}_j - \mu_j\|_{L_2(\mathbb{P})} = o_{\mathbb{P}}(1)$ or $\|\hat{\pi} - \pi\|_{L_2(\mathbb{P})} = o_{\mathbb{P}}(1)$ for consistency, and $\|\hat{\mu}_j - \mu_j\|_{L_2(\mathbb{P})} \times \|\hat{\pi} - \pi\|_{L_2(\mathbb{P})} = o_{\mathbb{P}}(n^{-1/2})$ for asymptotic normality. We therefore prefer $\widehat\psi^{(2)}_{\text{TMLE},\,j}$ for its weaker regularity conditions on nuisance estimation, and focus exclusively on this estimator for the rest of the manuscript. While one-step estimators that are consistent and normal under weaker conditions can be constructed (e.g., using the delta method and debiased estimators of the G-computation parameters) such an estimator is not guaranteed to be well defined, as the G-computation parameters may be estimated as negative, precluding a log transformation. See SI Section 8 for further discussion.

We now state sufficient conditions on the nuisance functions and their corresponding estimators that guarantee desirable asymptotic properties of our estimators of $\Psi^{(2)}(\mathbb{P})$. 
Although these conditions guarantee that both $\widehat\psi^{(2)}_{\text{OS}, j}$ and $\widehat\psi^{(2)}_{\text{TMLE},\,j}$ obtain normal limits, we prefer $\widehat\psi^{(2)}_{\text{TMLE},\,j}$ for its additional robustness properties in this setting. 

\begin{lemma} \label{result:nuisances}
    For every $j \in \{1,\dots,J\}$, assume that $X$ and $W$ are bounded and that $\pi$, $m_{j}$, and $q_j$ are cadlag functions with finite sectional variation norm. Suppose that $\hat{\pi}$, $\hat{m}_j$, and $\hat{q}_j$ are estimated using SuperLearner employing cross-fitting over a set of models that includes the Highly Adaptive Lasso.
    Then $\|\hat{\pi} - \pi\|_{L_2(\mathbb{P})} = o_p(n^{-1/4})$ and $\|\hat{m}_{j}\hat{q}_{j} - \mu_{j}\|_{L_2(\mathbb{P})} = o_p(n^{-1/4})$ for every $j \in \{1,\dots,J\}$.
\end{lemma}

A key advantage of SuperLearner \citep{SuperLearner} is that it is stable in finite samples (due to the inclusion of simple learners such as generalized linear models) while still guaranteeing the asymptotic convergence rates of complex learners (such as boosted trees and HAL \citep{HAL}). Note that the inclusion of fast, simple learners does not adversely affect the rates for nuisance estimation.

\begin{theorem} \label{result:efficiency3}
    Suppose $\mathbb{P}$ satisfies Assumptions \ref{assumption:mean_w}, \ref{assumption:joint_indep}, \ref{assumption:e_indep}, and \ref{assumption:s_joint_indep}.
    Assume further that the conditions of Lemma \ref{result:nuisances} are met and that nuisance functions are estimated as in Lemma \ref{result:nuisances}. 
    Then $\widehat\psi^{(2)}_{\text{TMLE}}$ is a regular and asymptotically linear estimator of $\Psi^{(2)}(\mathbb{P})$ with influence function $\phi^{(2)}(\mathbb{P})$. Therefore, $$\sqrt{n}\left(\widehat\psi^{(2)}_{\text{TMLE}} - \Psi^{(2)}(\mathbb{P})\right) \overset{d}{\rightarrow} \mathcal{N} \left( \textbf{0}_J, \Sigma^{(2)} \right)$$ for $\Sigma^{(2)} :=  \mathbb{E}_{\mathbb{P}}[\phi^{(2)}(\mathbb{P})(W,A,X) \times \phi^{(2)}(\mathbb{P})(W,A,X)^T]$. Thus, $\widehat\psi^{(2)}_{\text{TMLE}}$ is efficient in $\mathcal{M}$.
\end{theorem}

We now discuss the centered parameter $\Psi^{(2g)}(\mathbb{P})$, whose 
influence function is \begin{align*}
    & \phi^{(2g)}(\mathbb{P}) : (w,a,x) 
    \mapsto \begin{pmatrix}
    \mathbf{I}_{J} - \textbf{1}_J \times \nabla g \left(\Psi^{(2)}(\mathbb{P})\right)^T 
\end{pmatrix} \times \phi^{(2)}(\mathbb{P})(w,a,x)
\end{align*} where $\mathbf{I}_{J}$ is the $J$-dimensional identity matrix and $\textbf{1}_J \in \mathbb{R}^{J \times 1}$ is a vector of $1$'s. This parameter is a summary of the true category levels under different identifiability assumptions than Theorem \ref{result:efficiency3}. We apply the delta method and Theorem \ref{result:efficiency3} to obtain the following result about the estimator $\widehat\psi^{(2g)}_{\text{PI}} := \widehat\psi^{(2)}_{\text{TMLE}} - g\left( \widehat\psi^{(2)}_{\text{TMLE}} \right)$ of the centered parameter $\Psi^{(2g)}(\mathbb{P})$.

\begin{theorem} \label{result:efficiency4}
    Suppose $\mathbb{P}$ satisfies Assumptions \ref{assumption:mean_w}, \ref{assumption:joint_indep}, \ref{assumption:e_indep}, and \ref{assumption:s_cond_indep}.
    Assume further that the conditions of Lemma \ref{result:nuisances} are met and that nuisance functions are estimated as in Lemma \ref{result:nuisances}. 
    Then $\widehat\psi^{(2g)}_{\text{TMLE}}$ is a regular and asymptotically linear estimator of $\Psi^{(2g)}(\mathbb{P})$ with influence function $\phi^{(2g)}(\mathbb{P})$. Therefore, $$\sqrt{n}\left(\widehat\psi^{(2g)}_{\text{TMLE}} - \Psi^{(2g)}(\mathbb{P})\right) \overset{d}{\rightarrow} \mathcal{N} \left( \textbf{0}_J, \Sigma^{(2g)} \right)$$ for $\Sigma^{(2g)} :=  \mathbb{E}_{\mathbb{P}}[\phi^{(2g)}(\mathbb{P})(W,A,X) \times \phi^{(2g)}(\mathbb{P})(W,A,X)^T]$. Thus, $\widehat\psi^{(2g)}_{\text{TMLE}}$ is efficient in $\mathcal{M}$.
\end{theorem}

\subsection{Uncertainty quantification}

In the theorems above, we present that each of the estimators obtain normal limits with covariance matrix given by the covariance of the influence function. In order to perform statistical inference, we simply need to estimate that matrix. For example, to estimate the asymptotic variance of an efficient estimator of $\Psi^{(2)}(\mathbb{P})$, we can use the empirical analogue of \eqref{eq:variance_bound} using the estimated influence function: $ \widehat{\Sigma}^{(2)}(\widehat{\mathbb{P}}) := \mathbb{P}_n \left\{ \phi^{(2)}(\widehat{\mathbb{P}}) \phi^{(2)}(\widehat{\mathbb{P}})^T\right\}.$ 

As long as this estimator of the variance is consistent, we can construct Wald-type confidence intervals that have asymptotically valid coverage for our parameter. For example, an asymptotically valid $95\%$ confidence interval for $\Psi^{(2)}_j(\mathbb{P})$ is 
\begin{align}
    \left( \Psi^{(2)}_j(\widehat{\mathbb{P}}) - 1.96 \times \sqrt{\dfrac{\left[\widehat{\Sigma}^{(2)}(\widehat{\mathbb{P}})\right]_{jj}}{n}}, \quad \Psi^{(2)}_j(\widehat{\mathbb{P}}) + 1.96 \times \sqrt{\dfrac{\left[\widehat{\Sigma}^{(2)}(\widehat{\mathbb{P}})\right]_{jj}}{n}} \right).
\end{align}

In addition to marginal confidence intervals for each category, we can also construct simultaneous confidence intervals across $j\in\{1,\dots,J\}$ via an empirical single-step max-$T$ calibration approach \citep{Westfall.Young1993}. We draw $B$ bootstrap replicates $\hat{z}_b \overset{i.i.d.}{\sim} \mathcal{N}(\textbf{0}_J, \widehat{\mathrm{P}}_{\Psi^{(2)}})$, where $\widehat{\mathrm{P}}_{\Psi^{(2)}}$ denotes the empirical correlation matrix of the estimated influence function. The critical value for simultaneous confidence can be taken as the $1-\alpha$ quantile of $\{|\hat{z}_1|_\infty, \dots, |\hat{z}_B|_\infty\}$. 
Under weak correlation between the estimates, this approach will be no more efficient than applying a Bonferroni correction, but its efficiency relative to Bonferroni increases as the correlation increases. 


\section{Simulation study} \label{sec:sims}


We now evaluate estimation error and coverage of our proposed TMLE estimator and compare it to related methods. 
The best comparators for our approach come from the computational biology and statistical microbiome literature. 
Modern microbiome differential abundance methods acknowledge that the true levels of the categories $V$, which represent the total number of individual cells of a strain or species, are not observed. Instead, high throughput sequencing generates noisy observations $W$, which relate to $V$ through an assumption similar to Assumption \ref{assumption:mean_w} \citep{McLaren.etal2019}. While there are more than 30 methods for microbial differential abundance, we contrast our approach with the most widely used methods that target similar parameters to us: LinDA \citep{zhou2022linda}, ANCOM-BC2 \citep{lin2024multigroup}, MaAsLin3 \citep{nickols2024maaslin}, ALDEx2 \citep{fernandes2014unifying}, DESeq2 \citep{Love.etal2014}, and radEmu \citep{Clausen.Willis2024}. 
These methods all estimate a centered log-fold difference of the true abundance of bacteria between groups. 
We therefore believe $\Psi^{(2g)}(\mathbb{P})$ is the most comparable parameter to the estimand targeted by existing methods. 
We give a brief overview of all methods in SI Section 9, but 
the most similar comparison is with radEmu, which assumes log-linearity of $\mathbb{E}[W_{\cdot j} | A, X]$ in $A$ and $X$ and estimates target parameters via multinomial estimating equations (see SI Section 1.2 for the asymptotic bias of a related model). 
Note that \citet{Firth.Sammut2023} (discussed in Section \ref{sec:intro}) do not propose a specific methodology, and therefore no comparison is possible. 
We also do not contrast our method with approaches that target parameters on the proportion scale (e.g., \cite{Martin.etal2020}), do not accommodate covariate adjustment (e.g., \cite{brill2022testing}), or test an independence hypothesis without targeting a specific estimand (e.g., \cite{zhao2015testing, wang2023robust}).  
We refer to our method as ``niceday.'' 

We simulate data that reflects modern microbiome data: 
high variance; category-dependent sparsity; and non-independence of exposures and covariates. For each replicate of our simulation
we draw a continuous covariate $X \sim \text{Beta}(0.7, 1)$ and exposure $A |X \sim \text{Bernoulli}(0.95 \{ 
\arctan\left( 6 \times (X - 0.3) \right)/\pi + 0.5 \} + 0.05 ).$ For each category $j \in \{1, \ldots, J\},$ let $p_j$ be category-dependent sparsity, drawn as the $j$th element of a random permutation of $J$ equally spaced values on $[0.1, 0.9]$. We draw true category levels $V_j$ (true microbial abundances) from a zero-inflated negative binomial distribution: $V_j | A,X$ equals 0 with probability $1-p_j$ and is drawn from a $\text{NB}\left(\text{mean} = \gamma_j(A,X) / p_j, \;\; \text{size} = 2\right)$ with probability $p_j$. 
We consider two structures for the conditional mean $\gamma_j(a,x) :=  \mathbb{E}[V_j | A = a, X = x]$,
\begin{enumerate}[label=\Alph*]
    \item Log-linear: $\quad\;\; \gamma_{A,j}(a,x) := \exp\left\{ \left( 5 + \frac{1}{2}\log(j) \right) + a \times \left( 2 \log\left(\frac{2j}{J}\right) \right) - \left( \frac{1}{2} \log\left(\frac{j}{J}\right) \right) x \right\}$
    \item Log-nonlinear: $\gamma_{B,j}(a,x) := \exp\left\{ \left( 1 + \frac{5j}{J}\right) + a \times\exp\left(x \frac{2j}{J}\right) - \sin\left(\pi \left(x + \frac{j}{J}\right)\right) \right\}$. 
\end{enumerate}
Under both settings, $\gamma$ is monotone increasing in $j$ for fixed $a$ and $x$. The ordering of the categories $j$ is arbitrary, but a monotone mean simplifies visualization and analysis in a simulation study. 
We draw sample effects as $\text{Uniform}\left(0.1, 0.4\right) \text{ if } A=0$ and $\text{Uniform}\left(0.0,0.3\right) \text{ if } A=1$, and category-specific effects $E_j|A,X \sim \text{Gamma}\left( \text{shape} = 20, \; \text{rate} = 20 \times (J / (J+1-\tilde{j}))^{4/\log(J)} \right)$, 
where $\tilde{j}$ is the $j$th element of a random permutation of $\{1,\dots,J\}$. By construction, $\mathbb{E}_{\mathbb{P}}[S|A=1] / \mathbb{E}_{\mathbb{P}}[S|A=0] = 0.6$ (observations from cases are 0.6 times underrepresented compared to observations from controls, on average) and $\max_j \mathbb{E}_{\mathbb{P}}[E_j] / \min_j \mathbb{E}_{\mathbb{P}}[E_j] \approx 55$ (the most detectable category is 55 times overdetected compared to the least detectable category, on average). True category levels are unobserved, and the observed data are $W_j := V_j \times S \times E_j$. 
This data generating procedure satisfies Assumptions \ref{assumption:mean_w}, \ref{assumption:joint_indep}, \ref{assumption:e_indep}, and \ref{assumption:s_cond_indep}, which are required to identify $\Psi^{(2g)}(\mathbb{P})$.   
We fix $J = 51$, vary $n \in \{50, 100\}$, and perform 500 replicates for each simulation setting. True parameter values were calculated via numerical integration. Additional details on software versioning, hyperparameter selection, and nuisance estimation can be found in SI Section 9.  

\begin{figure}
\centering{
\includegraphics[width=0.8\textwidth]{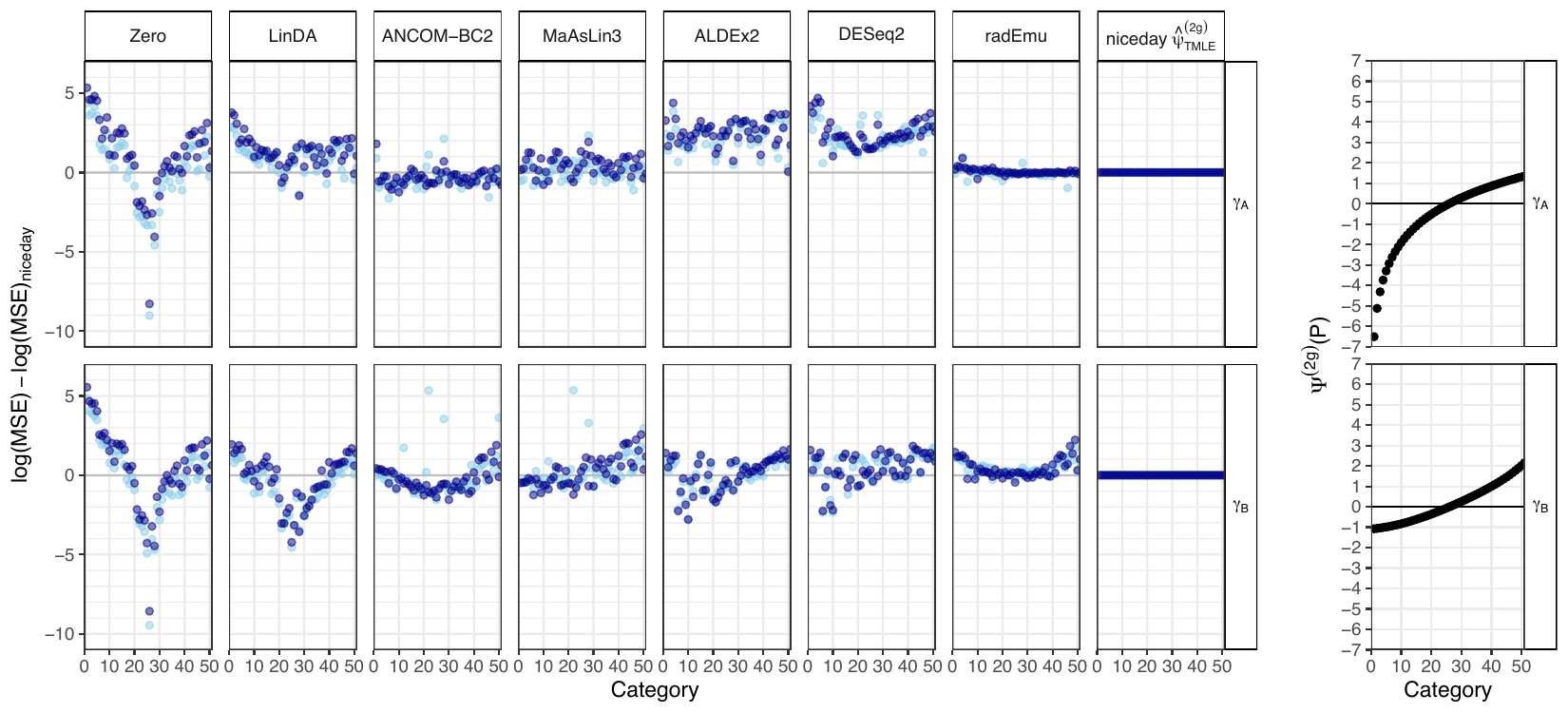}
}
\caption{log(MSE) of estimators for $\Psi_j^{(2g)}(\mathbb{P})$ compared to $\log \text{MSE}_{\text{niceday}}$, shown against the category index $j$. $\gamma_A$ corresponds to a log-linear true mean in both the contrast of interest $A$ and the adjustment covariate $X$, while $\gamma_B$ is nonlinear in $X$. True means are monotone increasing in $j$, thus estimators with large MSE at tail values of $j$ estimate large effect sizes poorly. No method uniformly outperforms niceday.}
\label{fig:mse_plot_ref}
\end{figure}

\textbf{Estimation error} Figure \ref{fig:mse_plot_ref} displays the log mean squared error (MSE) of all estimators compared to niceday (i.e., $\log \text{MSE}_{\text{estimator}} - \log \text{MSE}_{\text{niceday}}$); the equivalent visualization for $\log \text{MSE}_{\text{estimator}}$ can be found in SI Figure 1. We first see that no method outperforms niceday across all categories and simulation settings. 
Interestingly, some methods display category-specific patterns in their MSE: for example, LinDA estimates poorly for small and large $j$ but well for intermediate $j$. To understand this, we show the MSE of the estimator that always returns the estimate 0 (``Zero''; the MSE of this estimator is $\left(\Psi_j^{(2g)}\right)^2$). Similarities in the pattern in the MSE of Zero and a method indicates that the method performs particularly poorly in estimating large effect sizes. This can be the case in either all settings (LinDA, DESeq2) or in the nonlinear mean setting (ANCOM-BC2, MaAsLin3, ALDEx2, radEmu). The parameters of greatest scientific interest are those with the largest effect sizes, and therefore underestimating large effect sizes is a substantial practical disadvantage. niceday does not have this limitation (SI Figure 1). 


\begin{figure}
\centering{
\includegraphics[width=0.8\textwidth]{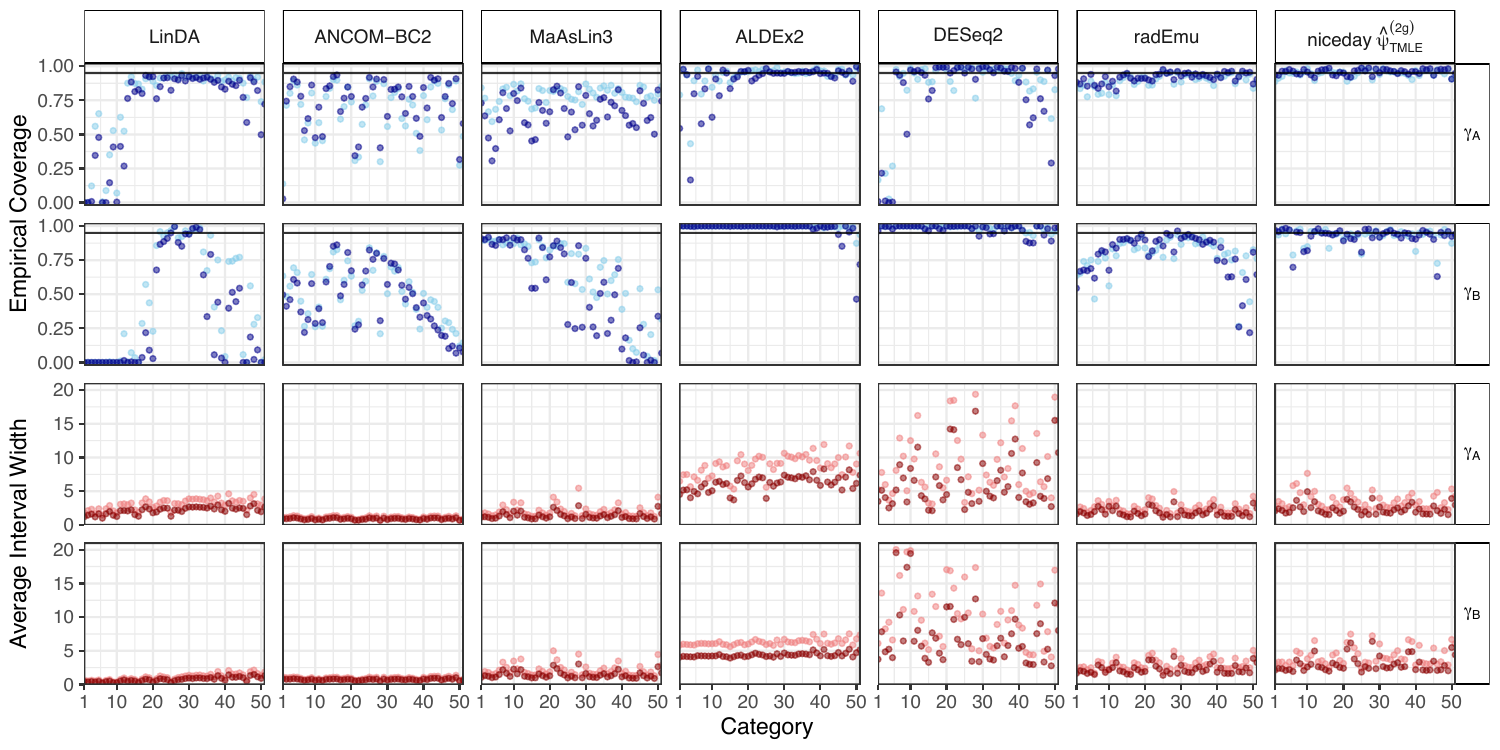}
}
\caption{Coverage of 95\% confidence intervals for $\Psi^{(2g)}(\mathbb{P})$, shown against the category index $j$. Nominal coverage is also shown (black line). 
The average width of confidence intervals is shown in the lower panel.
}
\label{fig:coverage_plot}
\end{figure}

\textbf{Interval coverage and width} In addition to point estimation, we also consider the validity and informativity of interval estimates. The empirical coverage of 95\% confidence intervals for $\Psi^{(2g)}(\mathbb{P})$ is shown in Figure \ref{fig:coverage_plot} (upper rows). LinDA, ANCOM-BC2, and MaAsLin3 suffer from under-coverage in all settings, while ALDEx2 and DESeq2 under-cover for log-linear means and over-cover for nonlinear means while the reverse is true for radEmu. These patterns can be explained by the average width of the confidence intervals (Figure \ref{fig:coverage_plot}, lower rows): LinDA, ANCOM-BC2, and MaAsLin3 produce the narrowest intervals while ALDEx2 and DESeq2 produce the widest. niceday, which consistently achieves nominal coverage, has slightly wider confidence intervals than radEmu. These results highlight the tradeoff of the flexibility of niceday: when the true mean is log-linear in the $a$ and $x$, niceday and radEmu both achieve nominal coverage, but niceday's intervals are wider. However, when the true mean is non-linear, niceday covers at the nominal level while radEmu undercovers. 

\textbf{Runtime} The median runtime for the non-linear mean $\gamma_B$ setting with $n=100$ for each method was (0, 6, 3, 14, 2, 2, 912) seconds for LinDA, ANCOM-BC2, MaAsLin3, ALDEx2, DESeq2, radEmu and niceday respectively. See SI Table 1 for runtime results across all simulation settings. Clearly, the estimation error and coverage guarantees of niceday induce a substantial computational cost. As is typically the case for debiased machine learning estimators like niceday, almost all of the computational cost arises from estimating the nuisances (here, 97\% of the runtime, on average). 
Simplifying the set of nuisance functions can reduce runtime at the expense of estimation error. 

\section{Data analysis} \label{sec:data_analysis}

We now illustrate our approach on a microbiome data application, investigating which bacterial strains are differentially abundant in populations with symptomatic diarrhea compared to those who are asymptomatic. 
We consider data from the EcoZUR (\textit{E. coli en Zonas Urbanas y Rurales}) study \citep{smith2019locals}. The EcoZUR study recruited patients in four regions of Northern Ecuador (Quito, Esmeraldas, Borb\'on, and rural communities outside of Borb\'on) between 2014 and 2015 using a case-control design. A case represents an individual presenting at a Ecuadorian Ministry of Public Health facility with diarrhea; upon admission of a case, a recently-presented age-matched individual from the same site without diarrhea was selected to be a control. We consider the 16S amplicon sequencing EcoZUR dataset \citep{jesser2023so}. After excluding participants with recent antibiotic exposure, we consider $n=352$ participants, of which 157 were diarrhea cases ($A=1$). We consider the following adjustment variables: age (in months), binarized sex, and recruitment site (indicator variables for each of the four sites). We consider $J=229$ strain-level categories of bacteria (16S sequence variants). 
We interpret all estimates relative to the smoothed median of fold-differences across bacterial strains (Result \ref{result:identifiability_2g}). 

\begin{figure}
\includegraphics[width=\textwidth]{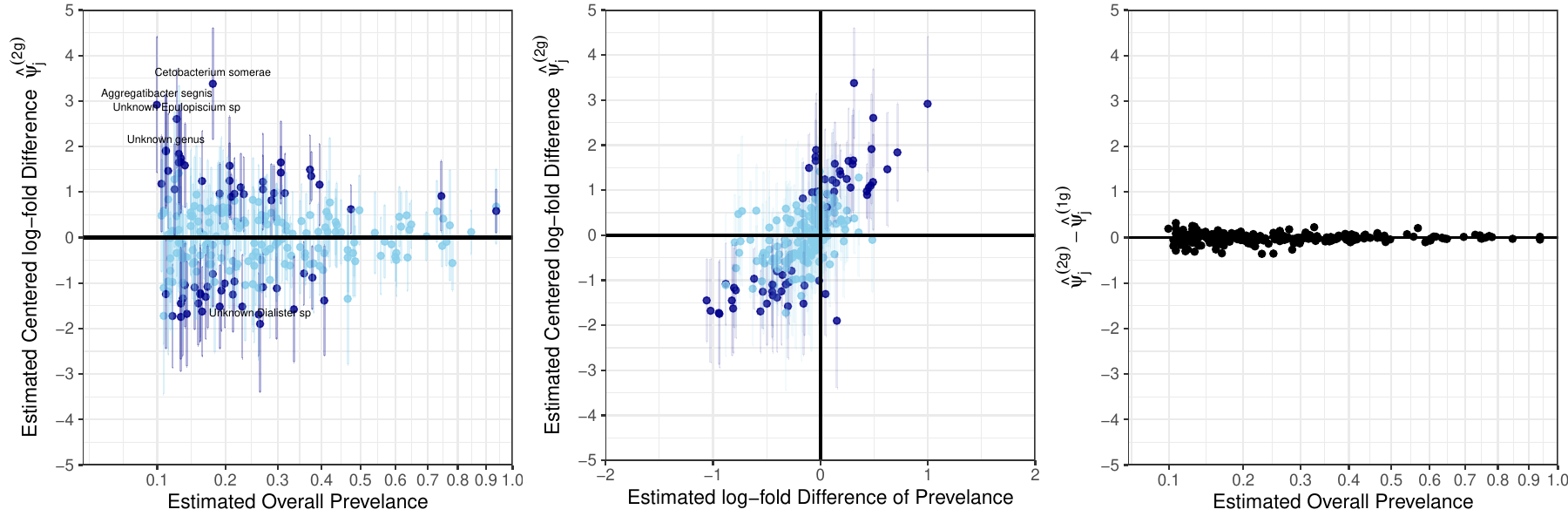}
\caption{
Estimates of adjusted log-fold differences in bacterial abundances across diarrhea cases and controls in a Ecuadorian cohort. Estimates $\widehat{\psi}^{(2g)}_{\text{TMLE},\,j}$ are shown against the overall prevalence of the bacterial strain in the study (left), and against the fold-difference in prevalence between cases and controls (center). Strains for which $H_0: \widehat{\psi}^{(2g)}_{\text{TMLE},\,j} = 0$ is rejected at $\alpha = 0.05$ are shown in dark blue. 
The difference between the adjusted and unadjusted estimators is also shown (right). }
\label{fig:DEC_estimates_niceday}
\end{figure}

Point estimates of log-fold differences in bacterial abundances across diarrhea cases and controls, marginalizing over age, sex and site and relative to the smoothed median log-fold difference, are shown in Figure \ref{fig:DEC_estimates_niceday} (left panel). Estimates $\widehat{\psi}^{(2g)}_{\text{TMLE},\,j}$ are plotted against the overall prevalence of the bacterial strain (percentage of samples in which the strain is detected), with dark colors indicating whether the marginal null hypothesis that the relative adjusted fold-difference equals the median across strains was rejected at the 5\% level. The strain estimated to be the most overabundant in diarrhea cases was \textit{Cetobacterium somerae}, with an estimated average abundance 29 times greater in cases compared to controls (95\% CI: 8.6 to 99.7). 
While this strain has not previously been associated with diarrhea, 
the genus \textit{Cetobacterium} has been associated with diarrhea in a Nicaraguan cohort \citep{Becker-Dreps.etal2015} and an unclassified strain of the genus was detected in a diarrheal fecal sample from Kolkata, India \citep{De.etal2020}. 
The strain estimated to be the second-most overabundant in diarrhea cases was \textit{Aggregatibacter segnis}, of the family Pasteurellaceae, with an estimated average abundance 18 times greater in cases compared to controls (95\% CI: 4.2 to 82.1).
This family was also detected in high abundance in diarrheal samples from Kolkata \citep{De.etal2020}. The third most differentially abundant strain belongs to the genus \textit{Epulopiscium}, also detected by \cite{De.etal2020}.
We finally note that the strain estimated to be the most overabundant in controls belongs to the genus \textit{Dialister}, estimated to be have an average abundance 7 times greater in controls compared to cases (95\% CI for $e^{-\Psi_{j'}^{(2g)}(\mathbb{P})}$: 1.5 to 29.7). In alignment, this genus has been previously associated with reduced symptom severity among patients with irritable bowel syndrome \citep{liu2020microbial}.


In Figure \ref{fig:DEC_estimates_niceday} (left panel), we see that significant strains occur across the full range of prevalence. We interrogate this further in Figure \ref{fig:DEC_estimates_niceday} (middle panel), where we show estimates against the fold difference in the prevalence of the strains in cases $\#\{i: A_i = 1 \text{ and } W_{ij} > 0\} /\#\{i: A_i = 1 \}$ compared to controls: $\#\{i: A_i = 0 \text{ and } W_{ij} > 0\} /\#\{i: A_i = 0 \}$. Unsurprisingly, strains that are more often present in cases than in controls tend to have large and positive estimated effect sizes. That said, strains with the same fold-differences in prevalence may have very different effect sizes. This reflects that the target parameter captures average presence-conditional abundance as well as prevalence patterns. 

We also compare our adjusted estimator $\widehat{\psi}^{(2g)}_{\text{TMLE}}$ to the unadjusted estimator $\widehat{\psi}^{(1g)}_{\text{PI}}$ in Figure \ref{fig:DEC_estimates_niceday} (right panel). Differences between these parameter estimates are negligible relative to their uncertainty. 
While the age-matching of the study design within site only guarantees that the distribution of age is the same for cases and controls within each site, the joint distribution of age, site and sex appears similar for both cases and controls (SI Figure 2). If $X \perp A$, then 
$\mathbb{E}_{X}[\mathbb{E}[V_j|A=1,X]] = \mathbb{E}_{X|A=1}[\mathbb{E}[V_j|A=1,X]]  = \mathbb{E}[V_j |A=1]$, where the last equality follows from the law of total expectation. Therefore, since $X \perp A$ implies $\Psi^{(1g)}(\mathbb{P}) = \Psi^{(2g)}(\mathbb{P})$, and $X \perp A$ holds approximately for this dataset, it is not surprising that $\widehat{\psi}^{(2g)}_{\text{TMLE}} \approx \widehat{\psi}^{(1g)}_{\text{PI}}$ here. 
This analysis does not suggest that these estimates will be similar in other applications. 
However, in this specific study, it aligns with our expectations and serves as validation for our approach. 

\begin{figure}
\centering{
\includegraphics[width=0.8\textwidth]{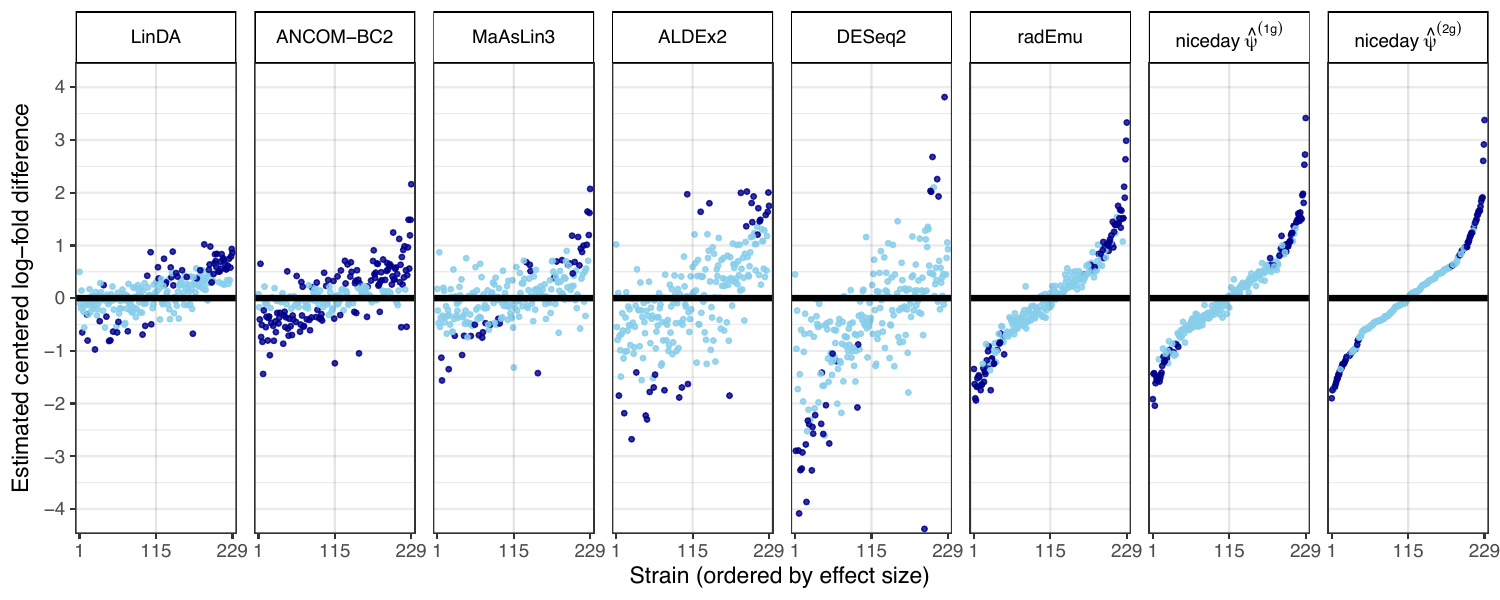}
}
\caption{Log-fold differences in true abundances across groups estimated using observed sequencing data via the proposed approach and comparable existing methods. Dark blue points indicate that the $95\%$ marginal confidence interval for that strain's parameter excludes zero. Estimates from radEmu are most similar to estimates from niceday, which is unsurprising given commonalities in their assumptions.}
\label{fig:DEC_estimates}
\end{figure}

We compare $\widehat{\psi}^{(2g)}_{\text{TMLE}}$ to comparable log-fold differences estimated by other methods in Figure \ref{fig:DEC_estimates}. Strains are shown on the $x$-axis in ascending order of the estimate $\widehat{\psi}^{(2g)}_j.$ 
As noted previously, $\widehat{\psi}^{(2g)}_{\text{TMLE}}$ and  $\widehat{\psi}^{(1g)}_{\text{PI}}$ yield similar estimates. We also see that the same is true for $\widehat{\psi}^{(2g)}_{\text{TMLE}}$ and radEmu. This is unsurprising for several reasons. As discussed previously, age is the only continuous variable for which we perform adjustment, yet age-matching strongly reduces any confounding that might be due to age. Further, the entire covariate distribution appears balanced between cases and controls, and so we would not expect major differences from the estimators with and without adjustment. Second, with the exception of the continuous variable of age, the log-linear mean assumption made by radEmu is trivially satisfied by binary indicators. Finally, niceday and radEmu impose similar identifiability assumptions, particularly, radEmu's mean model is highly similar to Assumption \ref{assumption:mean_w}. In contrast, the remaining methods show much less concordance, though effect sizes are positively correlated. $\widehat{\psi}^{(2g)}_{\text{TMLE}}$'s estimates are often larger than other methods. However, we showed in Section \ref{sec:sims} via simulation that LinDA, ANCOM-BC2, MaAsLin3, ALDEx2 and DESeq2 can underestimate large effect sizes, providing a plausible explanation for this pattern. 


\begin{figure}
\centering{
\includegraphics[width=0.7\textwidth]{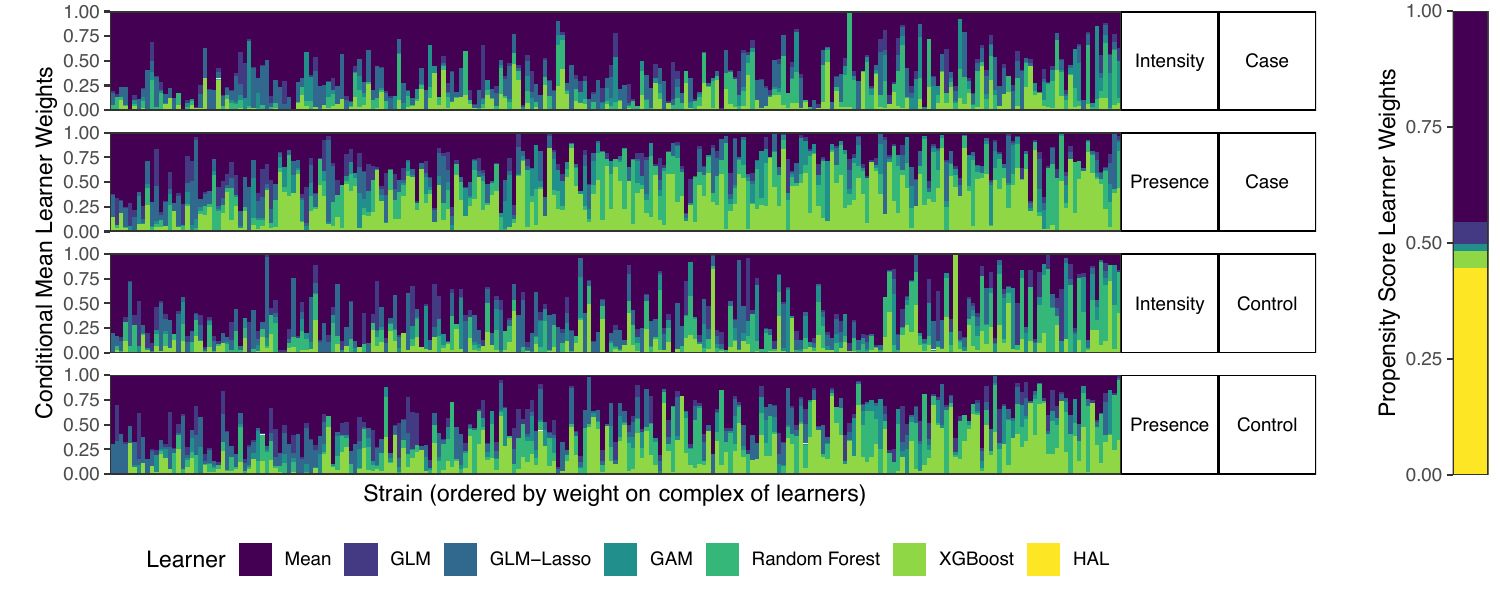}
}
\caption{The proposed method requires estimation of nuisance functions, for which we use a weighted ensemble of learners. We show the weights for these learners here. for both conditional mean (intensity and presence; left) and propensity score estimation (right). While these functions could be estimated using simple learners for many strains (e.g., sample mean), others required more complex learners (e.g., boosted trees). Averages of weights over cross-fit folds are shown. }
\label{fig:learner_weights}
\end{figure}

One advantage of niceday is that nuisance estimation can be performed flexibly using machine learning instead of assuming parametric (e.g., log-linear) models. To estimate nuisances, we use SuperLearner \citep{SuperLearner}, which performs an ensemble of regression procedures, weighting the best-performing learners using cross-validated risk minimization. Therefore, to understand the complexity of the nuisances, we can inspect the weights assigned to each learner. Figure \ref{fig:learner_weights} (left panels) show the weights of learners of the intensity $m_j(a,x) := \mathbb{E}[W_j|W_j>0,A=a,X=x]$ and presence $q_j(a,x) := \mathbb{P}(W_j>0|A=a,X=x)$ for both cases ($A=1$) and controls ($A=0$) across strains $j$ ordered by the weight assigned to flexible (versus simple) learners. We included the following flexible learners for the conditional mean: generalized additive model (GAM), random forest, and boosted trees; and the following simple learners: the sample mean, a generalized linear model with Poisson link (GLM) and a penalized GLM. In addition, we also included the highly adaptive lasso (HAL), a flexible but computationally expensive learner, for estimating the propensity score. 
The frequent selection of learners that are more complex than a penalized GLM suggests that a log-linear conditional mean $\mu_j(a,x)$ is not supported by these data. 
Figure \ref{fig:learner_weights} (right) shows that weights were assigned primarily to either the sample mean or HAL for propensity score estimation. 
If it were true that $A \perp X$, then $\pi(X)=\mathbb{P}(A=1|X)=\mathbb{P}(A=1)$, i.e., the sample mean is a good estimate of the propensity score for all values of the adjustment variables. 
However, the selection of HAL suggests that there may be meaningful imbalances in treatment assignment across covariate groups, which is unsurprising since matching was only performed by age and site.
Note that if the sample mean were always chosen for all learners, the estimator $\widehat{\psi}^{(2g)}_{\text{TMLE}}$ would be exactly equal to $\widehat{\psi}^{(1g)}_{\text{PI}}$; this provides another way to understand how covariate adjustment affects estimation. 


Finally, we investigated differences between the distributions of the sequencing depth ($\sum_j W_{ij}$) across $A$ and $X$. While identifiability assumptions regarding latent variables are inherently untestable, sizable differences in the sequencing depth suggest extra caution when applying Assumptions \ref{assumption:s_joint_indep} or \ref{assumption:s_cond_indep}. That said, we observe no differences in the distribution of sequencing depth across any stratification variables (SI Figure 3).


\section{Discussion}
\label{sec:discussion}

In this paper, we applied the toolkit of semi-parametric efficiency theory and Double Machine Learning to estimate ratios of means of categories when the observed data differs from the true category levels by both sample- and category-specific multiplicative terms. 
In contrast to many existing methods that address related questions (including Multinomial GLMs, and differential expression and differential abundance methods in the genomics literature), our approach is unique: instead of constructing an estimator via a family of parametric distributions or a generalized linear model, we began by specifying a parameter to target. We then stated assumptions needed to identify the parameter, and derived one-step and targeted minimum loss based estimators. The resulting estimators attain the smallest possible variance for consistent estimators over the class of distributions that satisfy the identifiability assumptions. This approach sidesteps the need to assess if a specific parametric model (e.g., overdispersed Binomial \citep{Martin.etal2020}, Negative Binomial \citep{Love.etal2014, Zhang.Yi2020, Jiang.etal2023}, Poisson \citep{Xu.etal2021}, multinomial \citep{Xia.etal2013, Zhang.Lin2019, Cao.etal2020, wang2023robust, Tang.Chen2019, Koslovsky2023}, or higher-order interaction models \citep{Willis.Martin2022,Scealy.Wood2023}) provides an adequate fit to any given dataset. Most importantly, our approach targets a clear and well-defined parameter, and does not make assumptions about the relationships between the average levels of the categories and covariates (e.g., log-linearity). 

We defined target parameters via means of the (unobserved) true category levels $V_j$, namely $\Psi_j^{(1)}(\mathbb{P}) = \log\left( \frac{\mathbb{E}\left[V_{ j}|A = 1\right] }{ \mathbb{E}\left[V_{ j}|A = 0\right]} \right)$, $\Psi_j^{(2)}(\mathbb{P}) = \log\left( \frac{\mathbb{E}\left[\mathbb{E}\left[V_{ j}|A = 1, X\right]\right]}{\mathbb{E}\left[\mathbb{E}\left[V_{ j}|A = 0, X\right]\right]} \right)$, and their centered counterparts. These parameters are well-defined as long as there is a non-zero probability of category $j$'s presence in both cases and controls: $\min\limits_a \mathbb{E}[V_{ j}|A = a] > 0$ for $\Psi_j^{(1)}(\mathbb{P})$ and $\min\limits_a \mathbb{E}[\mathbb{E}[V_{ j}|A = a, X]] > 0$ for $\Psi_j^{(2)}(\mathbb{P})$. In targeting these parameters, we do not require that the observed abundances $W_{ij}$ are positive for all $i$ and $j$, and therefore do not require common compositional data manipulations (``pseudocounts''). 
That said, alternative parameters to $\Psi_j^{(1)}(\mathbb{P})$ and $\Psi_j^{(2)}(\mathbb{P})$ exist. For example, we initially considered targeting $\Psi^{(3)}(\mathbb{P}) := \mathbb{E}\left[ \log\left( \frac{\mathbb{E}[V_j|A=1,X]}{\mathbb{E}[V_j|A=0,X]} \right)\right]$. 
A key advantage of this parameter is that $\Psi^{(3g)}(\mathbb{P}) := \Psi^{(3)}(\mathbb{P}) - g(\Psi^{(3)}(\mathbb{P}))$ can be identified under Assumptions 1, 2, and 3, whereas estimating $\Psi_j^{(2g)}(\mathbb{P})$ also requires Assumption 4.
However, $\Psi^{(3)}(\mathbb{P})$ is only defined if $\inf\limits_{a,x} \mathbb{E}_{\mathbb{P}}[V_j|A=a,X=x] > 0$ for all $j \in \{1,\dots,J\}$, whereas $\Psi^{(2)}(\mathbb{P})$ allows some subpopulations to be entirely absent of a category. Many applied scenarios exist where categories will be absent in some covariates (e.g., in ecology, the number of koalas in a forest in North America will always be zero). Therefore we believe that $\Psi^{(3)}(\mathbb{P})$ is of limited utility in practice. 
Furthermore, $\Psi^{(3)}(\mathbb{P})$ will be comparatively more sensitive to variations in $\mathbb{E}\left[V_{ j}|A = 1, X = x\right] / \mathbb{E}\left[V_{ j}|A = 0, X = x\right]$ across $x$, as this ratio may become large under the logarithmic transformation. In contrast, the iterated expectation of $\Psi^{(2)}(\mathbb{P})$ smooths these variations prior to the transformation. 
We leave the development of nonparametric estimators of $\Psi^{(3)}(\mathbb{P})$ to  future work, but give its efficient influence function in SI Section 4.

A major contribution of our work was to present sets of assumptions under which ratios of multi-category means are identifiable in the presence of category and sample distortion (Assumption \ref{assumption:mean_w}). While our assumptions place no restrictions on the form of the relationship between true category means and covariates, the independence assumptions may not hold in many scenarios. 
For example, in the microbiome setting, if samples were obtained in different batches, then Assumptions \ref{assumption:mean_w} and Assumption \ref{assumption:e_indep} are unlikely to be satisfied (treating batch as an adjustment variable). 
Furthermore, if any covariate is associated with the sample effects $S$ then Assumptions \ref{assumption:s_joint_indep} and \ref{assumption:s_cond_indep} will not be satisfied. This could be the case if, for example, DNA extraction is adversely affected by a more saline environment, and an adjustment variable (e.g., ocean temperature) is known to impact acidity (e.g., in coral microbiome samples). 
That said, we believe that these assumptions will generally be upheld in studies of relatively comparable samples -- for example, microbiome studies of the same sample type across interventions or diseases, but not cross-biome comparisons. While comparisons of broadly similar samples are common, we strongly encourage users of our method to consider the plausibility of the assumptions on a case-by-case, and only apply our estimators in settings where the assumptions are likely to be upheld. 
If the assumptions are not upheld, estimators may not be consistent, let alone efficient. 
That said, we view the clear statement of assumptions required by our method as an advantage, not a disadvantage, of our approach.

Our work admits many extensions. Firstly, parameters other than $\Psi^{(1)}_j(\mathbb{P})$ and $\Psi^{(2)}_j(\mathbb{P})$ could be considered. In particular, we focused on the setting of a binary exposure $A \in \{0,1\}$, but continuous exposures could also be considered. One approach to doing so is analogous to the framework of shift interventions. For example, under Assumptions \ref{assumption:mean_w}, \ref{assumption:joint_indep}, \ref{assumption:e_indep} and \ref{assumption:s_joint_indep}, we can identify 
\begin{align}
    \Psi^{(4)}(\mathbb{P}) & := \int \limits_{\mathcal{A}}{\dfrac{1}{\delta} \log\left( \dfrac{\mathbb{E}[\mathbb{E}[V_j | A=a+\frac{\delta}{2},X]]}{\mathbb{E}[\mathbb{E}[V_j | A=a-\frac{\delta}{2},X]]} \right)} \mathbf{1}\left(a - \tfrac{\delta}{2} \in \mathcal{A}, a + \tfrac{\delta}{2} \in \mathcal{A}\right) d \mathbb{P}_A(a),
\end{align}
and under Assumptions 
\ref{assumption:mean_w}, \ref{assumption:joint_indep}, \ref{assumption:e_indep} and \ref{assumption:s_cond_indep}, we can identify $\Psi^{(4)}(\mathbb{P}) - g(\Psi^{(4)}(\mathbb{P}))$. These parameters depend on a shift window $\delta$, but as $\delta \to 0$, this parameter describes the average percentage change in category levels for an increase in the exposure by one unit, marginalized across exposure values, $\int\limits_{\mathcal{A}} {\frac{d}{du}\log(\mathbb{E}[\mathbb{E}[W_j|A=u,X]]) \big|_{u=a}} d\mathbb{P}_A(a)$. If it were the case that $\log(\mathbb{E}[\mathbb{E}[V_j|A=a,X]]) = \gamma_j + \alpha_j a$, then $\Psi^{(4)}(\mathbb{P}) = \alpha_j$. In the general case, we can consider this parameter as the average slope of the true log-marginal mean. We leave development of this estimator and alternatives for future work.

Our work could also be extended to estimate causal equivalents of $\Psi^{(1)}_j(\mathbb{P})$ and $\Psi^{(2)}_j(\mathbb{P})$. Following standard causal notation, we let $V_{ j}(a)$ denote the level of category $j$ had the observational unit been assigned $A=a$. Under the standard causal inference assumptions \citep{robins1986new} of consistency ($A=a$ implies $V_{ j} = V_{ j}(a)$ for all $j$), no unmeasured confounding ($(V_{ j}(a) \perp A) | X$ for all $j$), and positivity (there exists $\delta > 0$ such that $1 - \delta \ge \mathbb{P}(A = 1|X = x) \ge \delta$ for every $x \in \mathcal{X}$), we have that $\Psi_j^{(1)}(\mathbb{P}) = \log\left( \frac{\mathbb{E}[V_{ j}(1)]}{\mathbb{E}[V_{ j}(0)]} \right)$ and $\Psi_j^{(2)}(\mathbb{P}) = \log\left( \frac{\mathbb{E}[V_{ j}(1)]}{\mathbb{E}[V_{ j}(0)]} \right)$. Specifically, if the treatment intervention $A$ is assigned randomly such that $(V(1), V(0)) \perp A$, the methodology described in this paper will target causal estimands that characterize how a treatment multiplicatively affects category levels. We leave further study of causal ratio estimation in the multi-category setting to future work. 

Finally, our work could be extended by imposing additional assumptions on the true or observed category levels. The advantages of imposing additional assumptions (beyond those required for identifiability) are that greater efficiency of estimation can be obtained, provided the assumptions hold. For example, we made no restrictions on the conditional mean $\mathbb{E}[V_j|A=a,X=x].$ However, if we believed this mean to be log-linear in the covariates $\log(\mathbb{E}[V_j|A=a,X=x]) = \alpha A +  \beta_0+ \beta^Tx$, we could derive an efficient estimator within the class of models satisfying this assumption. 
While our results show that log-linearity is not required for identifiability, we leave study of restricted estimators to future work. 

All estimators introduced in the main text are implemented in the R package \texttt{niceday}, 
available under a MIT license at \url{github.com/statdivlab/niceday}. Supporting Information (SI) for this paper will be provided from the corresponding author on request.

\section*{Acknowledgements}

This work was supported by NIH NIGMS R35 GM133420. 

\bibliographystyle{biometrika}

  \bibliography{niceday_bib}

\end{document}